\documentstyle{article}
\begin{document}
\begin{center}
{\bf 
 EXCHANGE COUPLING AND 
MAGNETIZATION PROFILES\\OF BINARY AND TERNARY MAGNETIC 
MULTILAYERS}\\[7pt]
 F.~S\"uss, U.~Krey\footnote{corresponding author} \\[5pt]
Institut f\"ur Physik II der Universit\"at, D-93040
Regensburg, Germany\\[5pt] and
S.~Krompiewski \\[5pt]
Institute of Molecular Physics, P A S \\
Smoluchowskiego 17, Pl-60-179 Pozna\'n, Poland 

\end{center} \begin{abstract}\noindent Within a spin-polarized LMTO
approach in the Atomic-Sphere approximation we calculate {\it ab
initio} the magnetic properties of various binary and ternary
multilayers composed of Fe, Co, Ni, Cr, V and Cu. The emphasis lies on
the indirect exchange interaction of the magnetic sandwiches across
the antiferromagnetic or non-magnetic spacers, and on the profiles of
the intrinsic resp.~induced magnetic moments. Among other results we
find: 
$(i)$ that Ni is very sensitive on its neighborhood and that $(ii)$
at the interface to Fe, V gets almost as strongly polarized as Cr,
whereas in the interior layers, the V sandwich remains
non-polarized. 
\end{abstract} \section{Introduction}

The exchange coupling between ferromagnetic metal sandwiches across
non-magnetic and antiferromagnetic spacers is a known interesting 
phenomenon, which is understood by various
theories using e.g.~the different reflection and interference
properties of electronic waves of different spin polarization in the
systems considered, e.g.~\cite{l:mathon,l:bruno}. Furthermore, it is
well known that the above-mentioned coupling depends in an oscillatory
way not only on the thickness of the spacer, but also on the
thicknesses of the magnetic layers of the system,
\cite{l:parkin,l:euro}, which offers additional possibilities to tune
the magnetic properties of the system. Finally, all these properties
are interesting with respect to possible applications, e.g.~by using
such systems as magneto-resistive reading-heads in magnetic storage
devices, since e.g.~the resistance strongly depends on the mutual
orientation of the magnetization of the ferromagnetic layers, which is
known as GMR-effect, see e.g.~\cite{l:GMR}.

To cope with the rapidly extending experimental possibilities to study
ternary and even more complex systems by using e.g.~double-wedge
techniques,\cite{l:bloemen}, we have extended our {\it ab-initio}
calculations to complex binary and ternary systems with up to 20
non-equivalent layers. Such {\it ab-initio} calculations have already
been used successfully by various groups,
e.g.~\cite{l:herman,l:dederichs,l:euro}, and although they typically
overestimate the amplitude of the thickness oscillations of the
exchange interaction by one order of magnitude compared to
experiments, the wavelengths and phases are reliably calculated, as
shown e.g.~by a comparison, performed in \cite{l:euro}, of results
from \cite{l:herman} with results obtained by the present method.

\section{Method and Results}

As in \cite{l:euro}, we use the first-principles LMTO package of
Anderson and Jepsen, \cite{l:anderson}, in its non-relativistic
version and in the Atomic-Sphere approximation, but with all important
corrections, by which: $(i)$ the accuracy of the k-space integrals is
enhanced 
 and $(ii)$ the overlapping and empty
parts of the elementary Wigner-Seitz-cell and -- at the same time --
higher angular momenta are taken into account.
 In the simplest case, the systems
considered are of the kind (n$_1$M$_1$/n$_2$M$_2$)$_\infty$, where
M$_1$ and M$_2$ are the magnetic or spacer metals, respectively, while
the n$_i$ denote the numbers of respective layers, and (...)$_\infty$
reminds to the periodic continuation of the antiferromagnetic unit
cell , which comprises 2(n$_1$+n$_2$) atoms, in all three
directions. The interface orientations considered are (0,0,1) and
(0,1,1), but below we only present results for the (0,0,1) case. The
exchange interaction $J$ is obtained from the energy difference
$\Delta E = E_{\uparrow , \downarrow} - E_{\uparrow ,\uparrow}$ for
the antiferromagnetic and the ferromagnetic orientation of this unit
cell by the equation 
 $J =\Delta E/(4A)$,
where $A$ is the area of the unit cell. Thus $\Delta
E > 0$ means ferromagnetic coupling.

In \cite{l:conicu} we have already studied the influence of Ni
replacing Co at the interfaces of a (Co$_4$/Cu$_n$)$_\infty$
multilayers and found a {\it weakening} of the interaction by almost a
factor $\sim$ 0.25, and a {\it phase shift} of the $\Delta E $-vs.-$n$ 
oscillations by almost one monolayer to the right for the
corresponding (Ni-Co$_2$Ni/Cu$_n$) systems.  Here we find that also
the magnetization of the Ni layers is {\it drastically influenced} by
its position with respect to the interface. A typical example is shown
in Fig.~1a: There the Ni moment is strongly reduced at the interface,
i.e.~from the bulk value 0.6 to $\sim 0.4 \,\,\mu_B$, whereas in the
central position of a Ni$_5$ sandwich one obtains a 10 \% enhancement,
and for the Ni$_7$ sandwiches even an oscillatory behaviour, with
enhanced values at layers 3 and 5, but a reduction almost to the bulk
value at the central layer 4. Another interesting observation is a
significant {\it induced polarization}, $\sim -0.015\,\, \mu_B$, of the
Cu layers at the interface.  Interestingly, this Cu polarization is
{\it antiparallel} to that of Ni, whereas it would be {\it parallel}
to Co at the corresponding Co/Cu interface, as we have found in
\cite{l:duesseldorf}.  Furthermore, we find for a ternary system
(NiCo$_n$NiCu$_2$)$_\infty$ in Fig.~1b that the induced Cu moments
mainly see the neighbouring Ni layer ond not the Co layers beyond.
Finally we found from systems with
$n\ge4$ Cu layers that the  induced Cu polarization is typically
reduced by a factor of $\sim 0.3$ at the 2nd Cu layer near the
interface, with respect to that of the 1st layer, i.e.~there is again a
significant polarization, and no change of sign, in layer 2. This should
be contrasted to the situation with a Co/Cu interface, where -- as
already mentioned -- the 1st Cu layer at the interface is polarized
antiparallel to Co ($\mu \approx -0.01\,\,\mu_B$), whereas the 2nd
layer must be polarized parallel to Co ($\mu \approx
+0.005\,\,\mu_B$), see \cite{l:duesseldorf}. 

In any case it should be noted that: $(i)$ such small induced Cu moments
of the order of 0.01 $\mu_B$ can be measured by X-ray circular
dichroism, \cite{l:schuetz}, and $(ii)$ that, according to our results,
Cu layers, which are {\it more} than two monolayers away from the
interface, are practically unpolarized.

Fig.2 shows that the {\it exchange} interaction
between the magnetic slabs in ternary multilayers of
Co, Ni and Cu changes drastically with the details of the
compositioni and the position of the various layers.
 E.g.\ for the Co-based multilayers of the lower curve
in Fig.\ 2, 
(Ni-Co$_4$-Ni/Cu$_2$)$_\infty$, (Co-Ni-Co$_2$-Ni-Co/Cu$_2$)$_\infty$,
(Co$_2$-Ni$_2$-Co$_2$/Cu$_2$)$_\infty$, (Co$_6$/Cu$_2$)$_\infty$,
 the exchange between 
the magnetic slabs is {\it antiferromagnetic}, i.e. $\Delta E < 0$:
 For the first system, with one Ni layer at the interface, it is
rather weak, but it becomes very strong  when the Ni layer moves into
the interior of the Co sandwich, i.e. for the 2nd and 3rd point
of the lower curve. In contrast, for the Ni-based 
multilayers of the upper curves, the exchange
{\it alternates} from antiferromagnetic ($\Delta E <0$)
 to ferromagnetic sign ($\Delta E > 0$) and 
{\it vice versa} in the sequence
(Co-Ni$_4$-Co/Cu$_2$)$_\infty$, (Ni-Co-Ni$_2$-Ci-Ni/Cu$_2$)$_\infty$,
(Ni$_2$-Co$_2$-Ni$_2$/Cu$_2$)$_\infty$, (Ni$_6$/Cu$_2$)$_\infty$. 
Generally speaking, the magnitude of the exchange is stronger when
Co, and not Ni, is at the interface.
 
Finally we have studied  (Fe$_n$/V$_m$)$_\infty$-multilayers. This
system is interesting on the following reason: V would be nonmagnetic
as a bulk metal. However, in the above-mentioned multilayers we have
found that the first V layer at the (001)-interface is strongly
polarized, $\mu =-0.6 \mu_B$, antiparallel to Fe. But in contrast
to Cr, all other V layers are non-polarized. Also the exchange
interaction $J$ of such systems shows strong oscillations as a
function of $n$ and $m$, see Fig.~3.

{\bf Acknowledgments}

This work has been carried out under 
the bilateral project DFG/PAN 436 POL and the KBN grant 
2P 302 005 07 (SK). 
Our thanks go also to the 
Munich, Pozna\'n and Regensburg Computer Centres.

\newpage
{\Large \bf Figure Captions}

Fig.~1a,b: Magnetization profiles in Ni/Cu and Co-Ni/Co multilayers.

Fig.~2: Exchange interaction in (Ni-Co$_4$-Ni/Cu$_2$)
 and related multilayers.

Fig.~3: Exchange interaction in (Fe$_2$/V$_n$) 
and
(Fe$_n$/V$_2$)
 multilayers.

\newpage 
\input epsf

\epsfxsize=13.5cm
\epsfbox{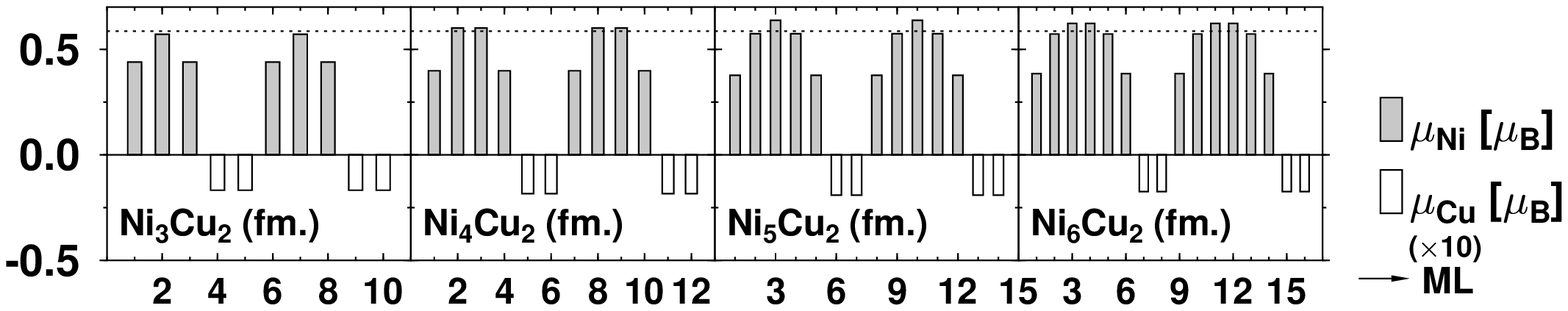}

\epsfxsize=13.5cm
\epsfbox{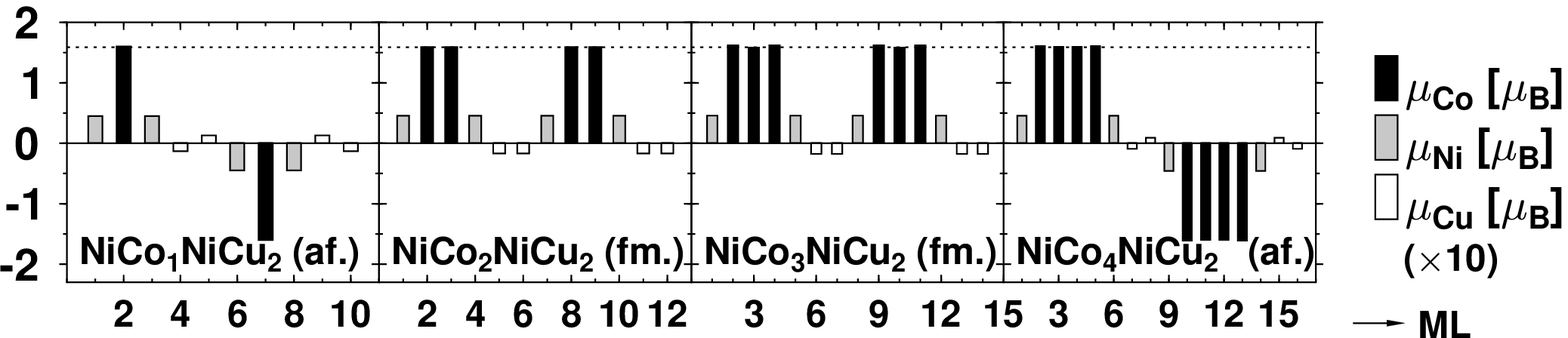}

Fig.1a,b: Magnetization profiles in Ni/Cu and Co-Ni/Co multilayers.

\newpage
\epsfxsize 12cm 
\epsfbox{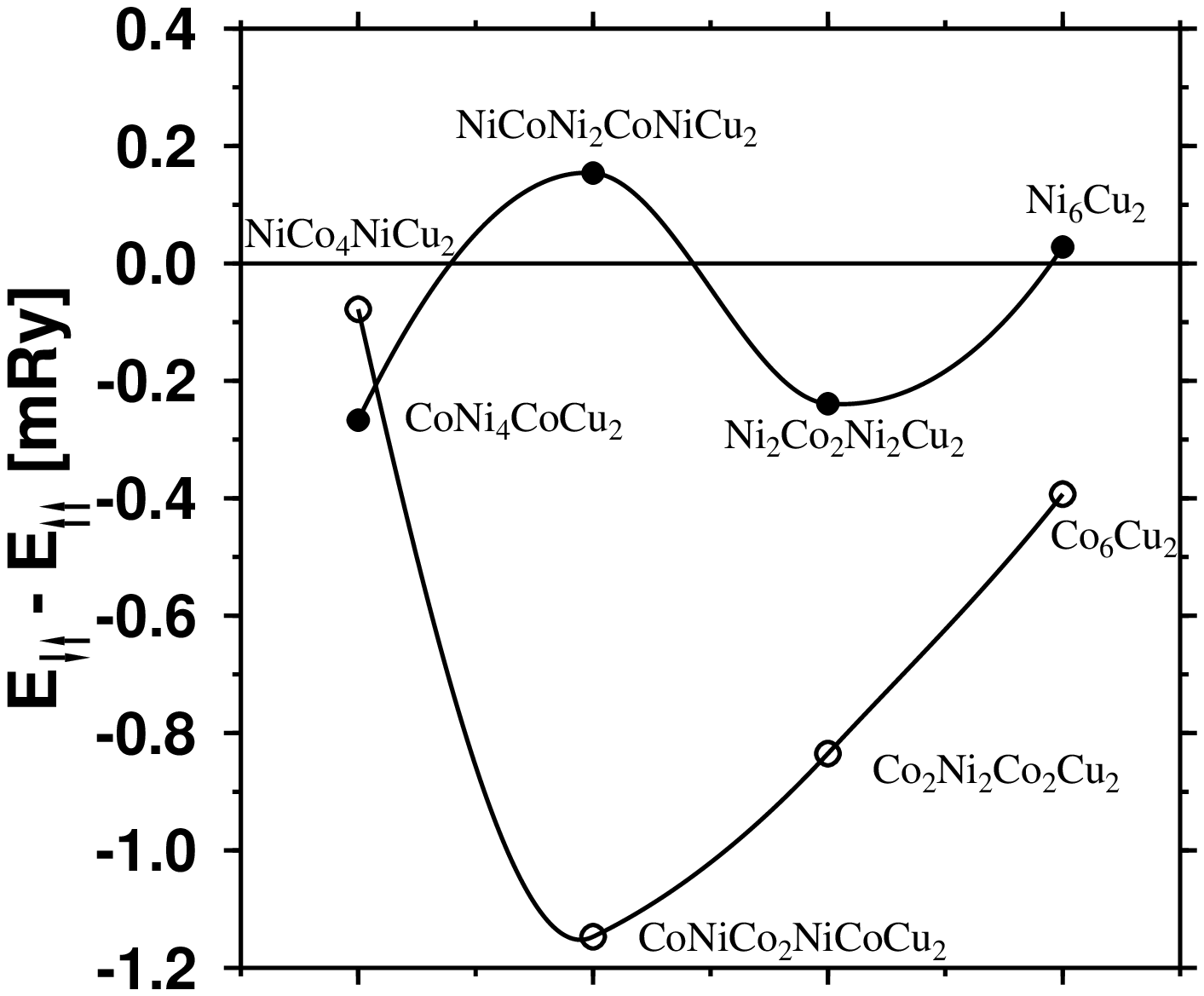}
Fig.2: Exchange interaction in (Ni-Co$_4$-Ni/Cu$_2$)
 and related multilayers.

\newpage
\epsfxsize=13cm 

\epsfbox{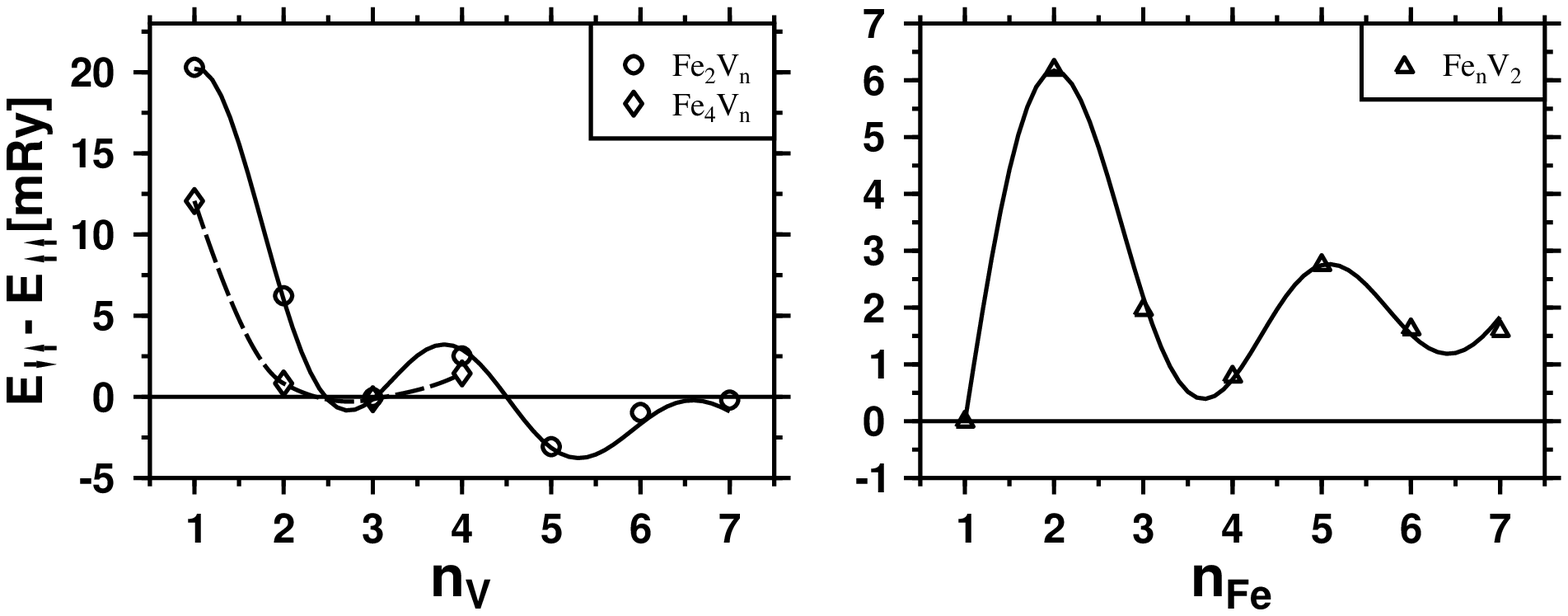}
Fig.3: Exchange interaction in (Fe$_2$/V$_n$) 
and
(Fe$_n$/V$_2$)
 multilayers.

\begin{thebibliography}{99}
\bibitem{l:mathon} D.M. Edwards, J. Mathon, R.B. Muniz, M.S. Phan,
{\it Phys. Rev. Lett.} {\bf 67}, 493 (1991).
\bibitem{l:bruno}  P. Bruno, {\it Phys. Rev. B} {\bf 52}, 411 (1995).
\bibitem{l:parkin} S.S.P. Parkin, N. More, K.P. Roche, {\it
Phys. Rev. Lett.} {\bf 64}, 2304 (1990).
\bibitem{l:euro} S. Krompiewski, F. S\"uss, U. Krey, {\it
Europhys. Lett.} {\bf 26}, 303 (1994).
\bibitem{l:GMR}M.N. Baibich, J.M. Broto, A. Fert, F.N. Van Dau,
 F. Petroff, P. Etienne, G. Creuzet, A.Friedel, J. Chazelas,
 {\it Phys. Rev. Lett.} {\bf 61}, 2472 (1988). 
\bibitem{l:bloemen}
P.J.H. Bloemen, M.T. van de Vorst, M.T. Johnson, R. Coehoorn,
W.J.M. de Jonge, {\it Mod. Phys. Lett. B} {\bf 9}, 1 (1994). 
\bibitem{l:herman} F. Herman, J. Sticht, M. van Schilfgaarde, {\it
J. Appl. Phys.} {\bf 69}, 4783 (1991).
\bibitem{l:dederichs} P. Lang, L. Nordstr\"om, R. Zeller,
P.H. Dederichs, {\it Phys. Rev. B} {\bf 50}, 13058 (1994).
\bibitem{l:anderson} We thank O.K. Andersen and O. Jepsen for 
their LMTO programs.
\bibitem{l:conicu} S. Krompiewski, F. S\"uss, U. Krey,
{\it J. Magn. Magn. Mater.} {\bf 149,} L251 (1995).
\bibitem{l:duesseldorf} S. Krompiewski, F, S\"uss, B. Zellermann,
U. Krey, {\it J. Magn. Magn. Mater.} {\bf 148}, 198 (1995).
\bibitem{l:schuetz} G. Sch\"utz, M. Kn\"ulle, H. Ebert, {\it Physica
Scripta} {\bf T49}, 302 (1993).
\end{thebibliography}
\end{document}